# Study of photothermal effect in core-shell transport microcontainers

Yury E. Geints [1]

[1] V.E. Zuev Institute of Atmospheric Optics SB RAS, Zuev square 1, 634021, Tomsk, Russia

E-mail: ygeints@iao.ru

**Abstract**

The theoretical study of laser-induced heating of a transport cargo microcontainer constituting a hollow dielectric microcapsule with water filling and a light absorbing polymer shell is presented. By means of FDTD and FEM techniques, the numerical calculations of optical field spatial distribution inside and near the microcapsules are carried out, and the temporal dynamics of temperature profiles of core-shell microparticles of various morphological types (spheres, ellipsoids, boxes, cubes, cylinders) and microcapsule clusters in different configurations is addressed. The spatial configurations of light absorption regions are shown to be substantially dependent on the morphology of the microcapsules. For isolated microcapsules, the most efficient laser heating is realized in capsules of cubic and cylindrical shapes. When the capsules are aggregated into the densely packed cluster, the characteristic heating depth of the whole assembly is higher for cylindrical microcapsules. In sparse microassemblies the square-shaped capsules are preferable for increasing the efficiency of laser heating.

## 1. Introduction

Miniature transport containers in the form of organic / inorganic hollow microparticles are the basis of the modern drug delivery systems, which can significantly increase the therapeutic effect while minimizing side effects due to the ultra-precise control of the spatiotemporal behavior of drug nanodoses inside the patient's body [1-3]. The controlled release of therapeutic molecules from microcapsules at the target places and at the right times becomes increasingly importance for advanced treatments including gene, antibody and vaccine therapies [4, 5].

The manufacturing technology of core-shell microparticles is a separate engineering task and to date can be implemented in various ways, from polymerization of monomers on emulsified micronuclei to poly-ionic layer-by-layer assembly (LbL) on solid-phase templates [6, 7]. A payload in the form of individual molecules or other active contents may be placed inside such a micro-container and then by blood or lymph flows it can deliver the cargo to the desired location where it can be activated or opened by an external stimulus. The external trigger may be, e.g., the change in the physicochemical properties of the environment (acidity, ionic concentration), some biological process (bacteria mediated degradation of the enzymes and polysaccharides in capsule shell), acoustic, magnetic or electromagnetic (optical) impulses [6].

The sensitivity of a capsule to the electromagnetic radiation is usually provided by doping the capsule shell with substances actively absorbing optical radiation in a certain spectral range. In the visible and near IR ranges, these can be noble metal nanoparticles, laser dyes, metal oxides [8-10]. The metal nanoparticles (NPs) deposited within the capsule shell transform the absorbed light into heat, which is released into the multicomposite material of the microcapsule matrix leading to an inhomogeneous distribution of its temperature [11]. This, in turn, can trigger or change the rate of photothermal reactions in the "hot areas" of the capsule shell and lead to its degradation and cargo release [12].

The capsules doped with metal NPs can exhibit absorption and scattering peaks at plasmonic resonances. The frequency and amplitude of plasmon resonance vibrations is influenced by the type of the constituting metal and also by a number of other factors such as the optical properties of NP surrounding, the degree of NPs aggregation, and their spatial shape [13, 14]. As shown in Ref. [15], the degree of light absorption by an initially transparent spherical dielectric microcapsule is effectively mediated by adding the specified amount of highly absorbing plasmonic gold NPs. with a rather low volume fraction of nano-inclusions in the capsule shell (~ 18%), it is possible to rise its absorption cross-section to the value of an absolutely absorbing sphere with a close to neutral absorption spectrum in the visible and near IR regions.

In most cases, the microcapsule is treated as a hollow micron-sized particle with a spherical shape. This spatial shape of capsules is most natural during their manufacturing process, e.g., by emulsion deposition method. However, the technology of LbL capsule assembly on solid nuclei (capsule fabrication by adsorption onto an organic or inorganic core) allows one to produce rather unusual types of micro-containers, which look like elongated geometric bodies such as ellipsoids, faceted cylindrical prisms, or cuboids [6, 16-18]. Recent study shows that the interaction potential of microcapsules of various forms with the targeted cells of the body may be also different [17].

Meanwhile, a typical method of delivering the required dose of cargo (medicine) is the accumulation of microcapsules in the target regions and subsequent massive opening of their shells by means of desired physical stimuli. In this case, the impact, e.g., of optical radiation will no longer be on a single microcapsule but on a cluster of such particles with different varying packing degree. This can drastically change the absorption characteristics of the microcapsules assembled in a cluster due to the optical fields interference from nearby particles, and consequently affect the efficiency of microcapsule shell laser heating [19, 20].

In this regard, a priority task is gaining the control on the heat release inside the microcapsule irradiated with an optical radiation. This is especially important because, on the one hand, there is a requirement in realizing the highest possible heat conversion efficiency of light energy. On the other hand, the heating of a transported substance (cargo) can be undesirable for its activation and further functioning. This requires the minimization of heating effect in certain areas of the particle. To address this challenge, the knowledges on the exact spatial configuration of the heat release regions and the temperature field inside the microcapsule exposed to a laser pulse are required.

In this paper, we address this issue and present the results of our theoretical studies of the absorption characteristics of optical radiation and the dynamics of heating of absorbing core-shell microcapsules. In the first part of the work, single microcapsules of several basic geometrical shapes are studied (spheroid, cuboid, cylinder) and the particles with the highest light absorption efficiency and heating ability are determined. In the rest of the paper the numerical solution of the heat transfer problem is carried out inside an ordered cluster of layered particles subjected to a pulsed optical radiation. For the first time to the best of our knowledge, the dynamics of heat-release regions and the maximal achievable temperature of microcapsules assembled in a cluster are studied in the dependence on the spatial arrangement and geometric shape of the constituent capsules.

## 2. Numerical model of a microcapsule

In the simulations, a microcapsule is modeled by a two-layer microparticle with a non-absorbing water core that simulates a payload and a solid-phase shell which can absorb optical radiation (Figure 1a). As the basis for a shell material the technical silicone is assumed doped with gold NPs of cylindrical shape (nanorods). According to our previous calculations, such a NP shape provides for the increased light absorption by the microcapsule in the near infrared optical band (Figure 1b) [15]. We consider the capsules of five geometric shapes: sphere, ellipsoid, cube, box, and circular cylinder. As an example, three types of microcapsules are shown in Figure 1c.

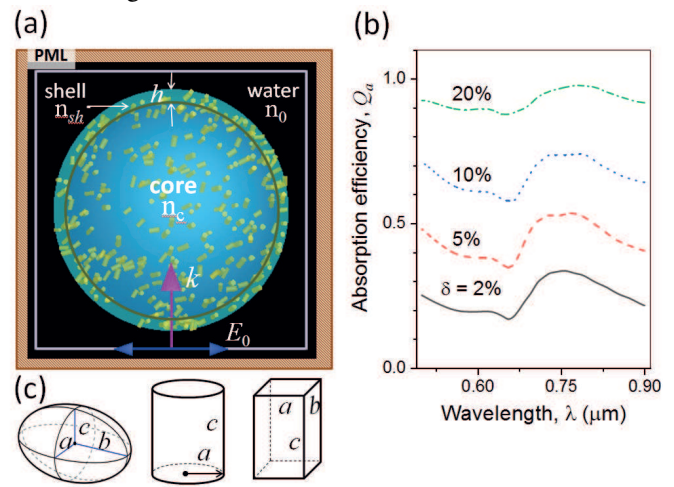

Figure 1. (a) Microcapsule model doped with light absorbing gold nanorods; (b) spectrum of light absorption efficiency $Q_a$ of a spherical capsule with different NPs volume content $\delta$; (c) basic morphological types of microcapsules.

Figure 1b shows the spectral behavior of light absorption efficiency by a spherical microcapsule obtained through the FDTD calculations. The absorption efficiency $Q_a = P_a/P_0$ is defined as the ratio of the optical powers being absorbed $P_a$ to the incident power $P_0$ on a particle. The diameter of the capsule is 1 μm with a shell thickness $h$ = 60 nm. This choice is supported by our recent studies showing that the maximal values of absorbed power density is achieved inside micron-sized capsules with a shell thickness of about 0.08λ that for λ = 800 nm gives the aforementioned value of $h$ [21].

As can be seen, the presence of nanorods inside the capsule shell (sizes 10×50 nm$^2$) leads to an absorption peak at λ = 0.755 μm, which corresponds to the excitation of the longitudinal surface plasmon resonance mode of such NPs. The spectral contour broadening of the plasmon resonance is explained by the random spatial orientation of the nanorods with respect to the plane of light wave polarization [15]. In this case, depending on the direction of polarization, a certain plasmon mode of a cylindrical NP is predominantly excited in the visible or IR spectral regions. A superposition of the contributions of both absorption channels of localized plasmon resonance with different intensities leads to the formation of a distorted Lorentz absorption contour with a blue-shifted maximum similarly to the Fano resonance excitation in a spatially asymmetric metastructure [22]. It is clear that the absorption efficiency of a microcapsule will be proportional to the volume content of the metal impurity.

For the specificity, in the calculations we set the NPs volume fraction in the capsule shell as δ = 10%. After the adaptation of the effective medium model (Bruggeman mixing rule), this yields a certain value of the complex refractive index $m_{sh} = n_{sh} - j\kappa_{sh}$ of the absorbing microcapsule. Thus, for the selected laser wavelength λ = 800 nm (fundamental harmonics of a titanium-sapphire laser), the effective optical parameters of the shell become: $n_{sh}$ = 3.6, $\kappa_{sh}$ = 0.07. The microcapsule core and the environment are considered as water with the refraction indices $n_0 = n_c = 1.33$ and zero absorption of infrared optical radiation.

The problem of absorbing electromagnetic radiation by a microparticle is solved by numerical integration of the Maxwell differential equations for field vectors in 3D-geometry. The equations are discretized using the finite-difference time domain method (FDTD) implemented in the commercial software package Lumerical FDTD Solutions. The computational domain is meshed by special cubic grid ("Yee cells") with staircase definition of magnetic and electric vectors. Mesh dimensions are set at $\lambda/(20|m|)$, where λ is incident radiation wavelength (in vacuum), and $m$ is medium complex refractive index, that results in about a million spatial cells with the sizes less than 10 nm. The time step of the numerical scheme is about 0.06 fs. The conditions of ideal field matching are applied at the boundaries of the computational domain, which are denoted as PML in Figure 1a. An adaptive anisotropic computational mesh is used thus providing nodes crowding within the regions of sharp medium permittivity gradients (layer boundaries of particle).

A plane linearly polarized wave with unit electric field amplitude $E_0$ and wave vector directed along the vertical axis (shown by arrows in Figure 1a) is assumed to illuminate the lower domain boundary. Light absorption efficiency $Q_a$ of a microcapsule is calculated based on the electric field distribution $\mathbf{E}(\mathbf{r})$ in the microparticle volume $V_c$ and the value of the total absorbed power (TAP) $P_a$ according to the following expression:

$$P_a = \frac{\pi c \varepsilon_0}{\lambda} \int_{V_c} d\mathbf{r}\, \varepsilon''(\mathbf{r}) |\mathbf{E}(\mathbf{r})|^2 = P_0 \int_{V_c} d\mathbf{r}\, q_a(\mathbf{r}) \quad (1)$$

Here, $\varepsilon_0$ is dielectric permittivity of vacuum, $\varepsilon''(\mathbf{r}) = 2n\kappa$ is the imaginary part of particle complex dielectric permittivity ($\varepsilon = \varepsilon' - j\varepsilon'' = (n - j\kappa)^2$), $c$ is the speed of light (in vacuum), $q_a(\mathbf{r}) = n\kappa B(\mathbf{r})/\lambda S$ is reduced absorbed power density (APD), $B(\mathbf{r}) = |\mathbf{E}|^2/E_0^2$ is the field enhancement factor, and $S$ stands for the midsection of the capsule depending on its spatial orientation relative to the optical wave incidence. As seen, the value $P_a$ is mainly influenced by the specific spatial distribution of optical field in the microcapsule, which is accounted by the field enhancement factor $B(\mathbf{r})$.

## 3. Shape-mediated light absorption by microcapsules

Consider the results of the numerical simulations on light absorption by microcapsules depending on their morphology. Figures 2(a), 2(c) show the spatial profiles of absorbed optical power taken in the cross section of microcapsules of various shapes. In each 2D-plot, the distributions of relative intensity $B$ and absorbed energy density $q_a$ are superimposed for visual aids. The capsules are assumed to have equal volumes $V_c = 10^{-15}$ liter regardless of their particular shape.

From these figures is clear that in all types of capsules considered the regions of principal light absorption ("hot areas") are located both in the illuminated and shadow parts. The latter is caused by the focusing effect of incoming optical wave by a mesoscale dielectric object. The absorbing shell of capsule does not prevent optical wave penetration into particle volume and the formation of diffraction maxima near the surface because the characteristic absorption scale $l_a = \lambda/4\pi\kappa_{sh}$ equals to 900 nm in this case that significantly exceeds the capsule shell thickness $h$.

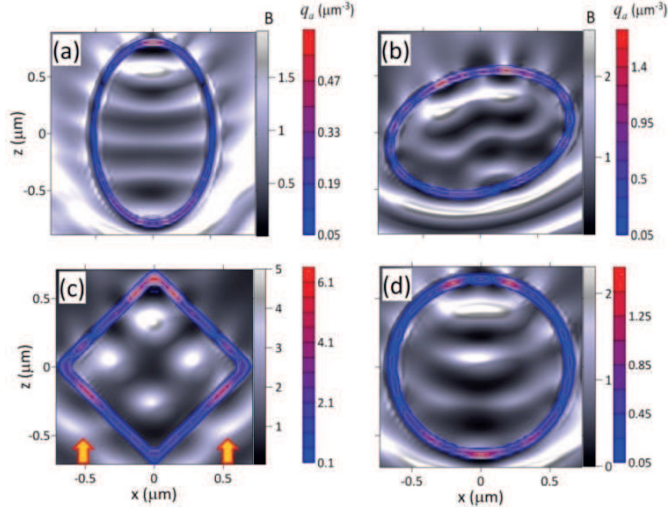

**Figure 2**. Spatial 2D-distribution of relative optical intensity $B$ (shaded color relief) and reduced absorbed power density $q_a$ (colored contours) in microcapsules with (a, b) elliptical, (c) cubic and (d) spherical cross section. Light incidence is shown by arrows.

At the same time, the capsule morphology introduces specificity in the character of optical field distribution and accordingly in the spatial position and extension of light absorption regions. Assuming the constant volume of a microcapsule, any deformations in its geometric shape can rebuild the optical intensity profile. According to Equation (1), two factors influence the value of the attainable absorption power density: field intensity enhancement $B$ inside the particle and shell absorption $\varepsilon''$. If these parameters in coordination reach their maximum values within the same area(s) of capsule, then in this region the absolute maximum of optical power absorption is realized. For particles of ellipsoidal shapes (Figures 2a, 2b, 2d), this situation is observed mainly near the shadow surface (along the radiation propagation), whereas in the cubic capsule (Figure 2c) a noticeable increase in the optical intensity occurs near all its vertices. Thus, among the considered types of microcapsules the cubic form gives the highest values of absorbed power density reaching $q_a \approx 6.1$ μm$^{-3}$.

The spatial orientation of the capsule with respect to the direction of optical radiation propagation and its polarization vector also plays a significant role in controlling the absorption profile. Consider Figures 2a and 2b, where the distributions of absorption regions inside an ellipsoidal capsule are shown differently oriented by its major semi-axis along the direction of light incidence; in Figure 2a the elliptic particle is oriented along the wave vector, and in Figure 2b it is rotated about an *y*-axis by an angle $\varphi_y = 65°$. By comparison of these figures one can see that in the latter case, a redistribution of the internal optical field is observed leading to the extension of spatial regions with enhanced intensity and to a more than threefold increase in the absorption power density.

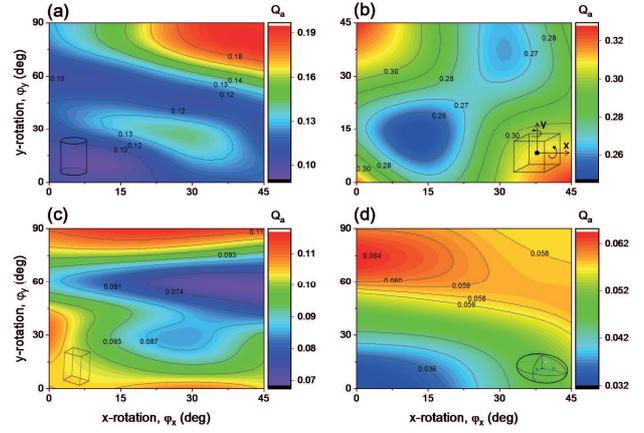

**Figure 3**. Orientation-dependent absorption efficiency $Q_a$ of microcapsules with various shapes: (a) cylinder, (b) cube, (c) rectangular box, (d) sphere.

Figures 3a-d show the dependence of total relative power $Q_a$ absorbed by various microcapsules on the orientation angles along the two coordinate axes $\varphi_x$ and $\varphi_y$. The case with $\varphi_x = \varphi_y = 0$ corresponds to the orientation of the particle with the elongated part in the direction of optical wave incidence. From these figures it follows that for each type of microcapsule there is a certain range of rotation angles that provide maximum absorbed power. In general, the absorption efficiency is maximum at the maximum particle midship, i.e. when $\varphi_y = 45°$ ($\varphi_x = 0$) for the cube and $\varphi_y = 90°$ for all other types of capsules. Moreover, the range of $Q_a$ variation caused by the particle rotation is also approximately the same for all considered shapes and amounts to a twofold change in magnitude. The only exception is a cubic capsule, where this variation is considerably less.

In real situations, however, it is challenging to realize any specific or desired spatial orientation of the microcapsule. This is why it is more reasonable to consider the effective absorption parameters ($Q_a$, $q_a$) averaged over all rotation angles. A comparison of various types of microcapsules based on their averaged absorption characteristics is shown in Figure 4. Here, for all spatial forms of particles, the values of maximal intensity enhancement $B_{max}$, reduced absorption power density $q_a^{max}$, and absorption efficiency $Q_a$ are presented. Note that during the averaging procedure, all particle orientations are considered as equiprobable.

As can be seen from the figure, the cubic capsule is the absolute leader in all respects. Even with the cube edge of only 1 μm (1.25λ), such a particle provides efficient focusing of incident optical wave when the internal field intensity increases by almost ten times. Importantly, that the loci of these intensity extrema are near the absorbing shell of the coated cube that ensures the realization of high absorption power density because the maxima of $B$ and $\varepsilon''$ distributions coincide.

The nearest competitors of the cubic capsule are cylindrical and rectangular particles, which demonstrate multiple lower values of absorbed power and optical intensity. And at the end of the list are ellipsoidal capsules possessing the lowest light absorption efficiency $Q_a$.

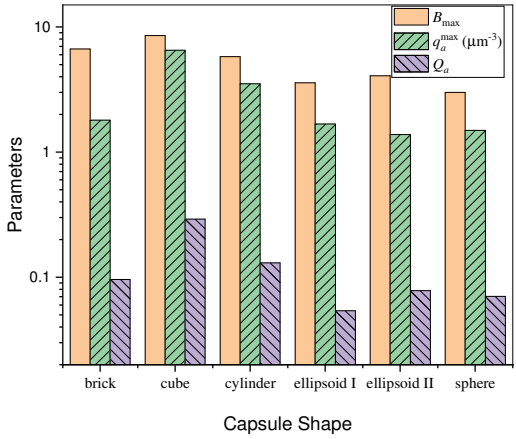

**Figure 4**. Orientation-averaged absorption parameters for different microcapsules.

Notice, in Figure 4, the morphological types of capsules denoted as "*ellipsoid I*" and "*ellipsoid II*" differ in the degree of elongation and correspond to the eccentricity decrease of elliptical cross section. Interestingly, with a low absorption of entire elliptical microcapsule the degree of internal optical field enhancement and the absorption power density in the shell are sufficiently high. This indicates that in an ellipsoid due to its curved surface, a sharp light wave focusing can be realized. Here, the regions where the absorption mainly takes place are small but the absorption density values are high.

## 4. Laser-induced heating of core-shell microcapsules

In this section, we discuss the heating dynamics of the microcapsules with a laser radiation. The corresponding thermophysical problem is considered within the isobaric approximation provided by the following inequality: $t^* \gg R/c_s$. Here, $t^*$ is the characteristic time scale of the problem, e.g., the duration $t_p$ of a laser pulse, $R$ is the characteristic particle size, and $c_s$ is the speed of sound in the medium. In this case, a fairly quick (on the pulse scale) pressure equalization occurs in cold and heated regions of medium and instead of considering the closed system of thermo-hydrodynamic equations, only single equation can be solved for time-dependent medium temperature $T(\mathbf{r},t)$ [23]:

$$\rho\, C_p \frac{\partial T(\mathbf{r},t)}{\partial t} = \lambda_T \nabla^2 T(\mathbf{r},t) + Q(\mathbf{r},t) . \qquad (2)$$

Here $\nabla^2$ denotes the Laplacian in three-dimensional space, $\rho$, $C_p$, $\lambda_T$ are the density, specific heat capacity at constant pressure and thermal conductivity coefficient, respectively, $Q(\mathbf{r},t)$ is the volumetric power density of heat sources due to absorption of light radiation. All thermodynamic characteristics are considered functionally dependent on the coordinate $\mathbf{r}$ of the point in the medium and its temperature $T$. Hereinafter, the temperature is understood as its change from the initial value $T_0 = 293$ K. In Equation (2), the convective heat fluxes between differently heated regions of medium are omitted that is valid when considering heat transfer processes on a microscale. In addition, it is assumed that medium advection near a particle is also absent and at the external boundaries of the computational domain $\mathbf{r} \in \mathbf{R}$, heat transfer is zero: $\nabla T\big|_{\mathbf{r}=\mathbf{R}} = 0$.

In Equation (2), the power density of heat sources $Q(\mathbf{r},t)$ depends on the spatio-temporal profile of the light intensity $I(\mathbf{r},t)$ inside a microcapsule: $Q(\mathbf{r},t) = \alpha(\mathbf{r}) I(\mathbf{r},t)$, where $\alpha = 4\pi\kappa/\lambda$ is the volumetric absorption coefficient of capsule substance. Generally, to obtain the optical field distribution inside a multilayer particle, it is necessary to solve the time-dependent wave equation accounting for the temporal variability of the electromagnetic field upon particle heating. However, this problem can be significantly simplified if we do not consider the transient processes of optical field distribution establishing in the vicinity of the particle, and rather consider the optical intensity as the quasi-stationary distribution with a spatial profile, which amplitude follows the time profile of the laser pulse.

Thus, once the thermodynamic equilibrium is established the distribution of the optical field in medium can be considered stationary and its spatial distribution is obtained by the Helmholtz equation:

$$\nabla \times \nabla \times \mathbf{E}(\mathbf{r}) - k^2 \varepsilon(\mathbf{r}) \mathbf{E}(\mathbf{r}) = 0 . \qquad (3)$$

Here, $\varepsilon$ is the dielectric permittivity of medium, and $k = 2\pi/\lambda$ is the optical wavenumber. The electric field **E** is a vector with the components along each of the coordinate axes. Worthwhile noting in the case of a multilayer spherical particle there is an analytical solution to Equation (3) in the form of infinite series of the vector spherical harmonics, called in the literature the "Mie series" [24]. For capsules of nonspherical spatial forms, such partial-wave expansion is too sophisticated (T-matrix method) or time-memory consuming (Discrete-Dipole Approximation) [25, 26]. Thus, we used the direct numerical integration of the wave equation (3) to derive the electromagnetic field distribution.

Numerical integration of Equations (2)-(3) is carried out using the COMSOL Multiphysics software, which implements the solution of differential equations by the finite element method (FEM). To define the thermophysical properties of microcapsule shell that in fact is constituted by two physically diverse substances (silicone and Au NPs), we use the COMSOL "Porous Medium" program node, which takes into account the distribution of heat in a two-component medium consisting of a porous matrix and a coolant. This procedure introduces a specific homogeneous medium with effective physical parameters calculated by the following models:

− Density:
$$(\rho)_e = \delta(\rho)_d + (1-\delta)(\rho)_m ; \quad (4)$$

− Specific heat capacity:
$$(\rho C_p)_e = \delta(\rho C_p)_d + (1-\delta)(\rho C_p)_m ; \quad (5)$$

− Heat conductivity:
$$(1/\lambda_T)_e = \delta(1/\lambda_T)_d + (1-\delta)(1/\lambda_T)_m \quad (6)$$

Here, the subscripts "e", "d", "m" denote the values of the effective (volumetric) and the corresponding values for the impurity (gold) and matrix (silicone) parameters. The effective thermodynamic and optical parameters calculated for each capsule layer are shown in Table 1.

Table 1. Physical parameters of capsule constituents ($T_0$ = 293 K, $\delta$ =0.1).

| Parameter Name | Core/Surrounding (H$_2$O) | Shell (Silicone+Au NPs) |
|---|---|---|
| $\rho$ [g/cm$^3$] | 1 | 3.07 |
| $C_p$ [kJ/kg·K] | 4.2 | 1.63 |
| $\lambda_T$ [W/m·K] | 0.6 | 0.29 |
| $\chi$ [µm$^2$/µs] * | 0.14 | 0.06 |
| $\alpha = 4\pi\kappa/\lambda$ [1/µm] | 0 | 1.1 |
| $n$ | 1.33 | 3.6 |

*$\chi = \lambda_T / \rho C_p$ is the heat diffusivity.

A rectangular laser pulse with typical parameters of femtosecond Ti:Sapphire lasers ($\lambda$= 0.8 µm, duration $t_p$ = 0.1 ps) is directed to the microcapsule. The total pulse energy is always constant and equals to 20 µJ. Figures 5(a, b) show the temperature distribution in the cross sections of cylindrical and ellipsoidal microcapsules at a time moment corresponding to the end of the laser pulse. The relative optical intensity profile is plotted in each figure along with the temperature distribution.

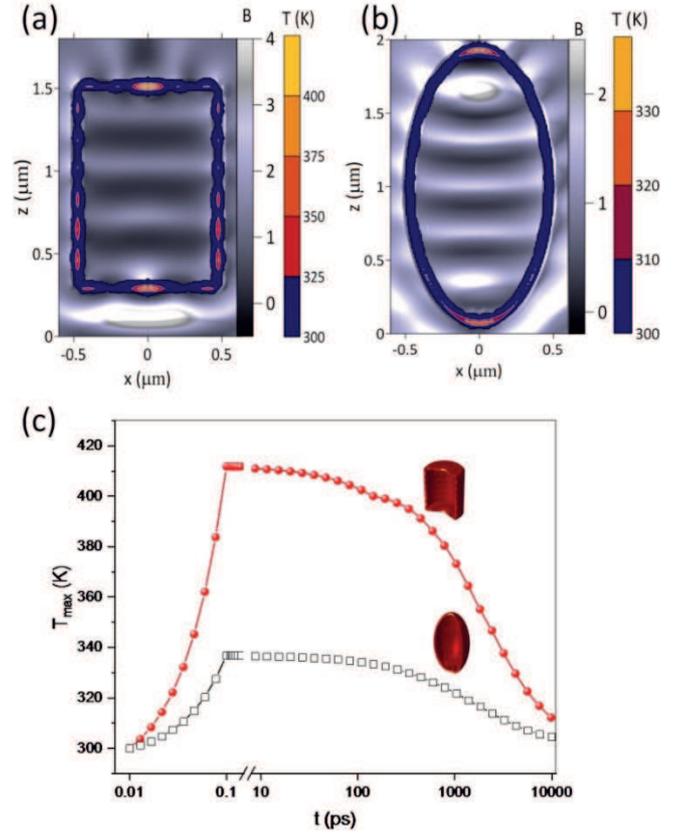

**Figure 5**. (a,b) Optical intensity *B* (shaded color relief) and temperature *T* (contours) 2D-profiles inside (a) cylindrical and (b) ellipsoidal capsules; (c) maximal temperature $T_{max}$ in the capsule shell versus time *t*.

It is seen from these figures that with such sufficiently short pulsed irradiation of hollow particle ($t_p \ll t_T = h^2/4\chi$ ~ 10 ns) its temperature distribution closely resembles the "hot areas" intensity profile. Since the light absorption occurs only in the shell, the regions of predominant capsule heating are concentrated in this layer with maxima along the laser beam incidence. Other parts of the absorption layer are also heated but their temperature is several times lower. The water core of the microcapsule and adjacent outer layers of environmental water remain cold.

The dependence of maximum temperature $T_{max}$ in various capsules is shown in Figure 5c. Two physical processes corresponding to different time scales are shown here. The first process is laser heating of a core-shell capsule during the pulse time of 0.1 ps; the second one describes the cooling of an unevenly heated particle due to heat diffusion from the heated shell into the cold capsule core and the environment. At this temporal stage the laser radiation is off.

Notice worthy, the maximum attainable temperature $T_{max}$ of a capsule is influenced by two main factors: (a) the laser pulse energy and (b) the particle morphology, i.e., its shape and internal structure. An increase in laser pulse power is always followed by an increase of capsule temperature, because from Equation (2) in the instant source approximation one has: $T_{max} \propto \left(I_0 t_p\right) B_{max}$, where $B_{max}$ is the maximum value of the optical field enhancement factor in the "hot areas". The parameter $B_{max}$ involves the dependence of heating degree on the specific design of a capsule. As follows from Figures 4 and 5c, even with a fixed capsule volume and the thickness of its absorbing layer a change in the geometric shape dramatically affects the dynamics of the heat release. Thus, the peak temperature of the shell of a cylindrical capsule reaches $T_{max}$ ~ 415 K towards the end of laser irradiation that is close to the threshold temperature of thermal destruction of silicone (420 K), which the capsule shell is made of. Under the same conditions, the capsule in the spatial form of an ellipsoid is heated to a maximum temperature $T_{max}$ ~340 K. Clearly, this fact must be taken into account when designing transport microcontainers sensitive to laser radiation.

## 5. Laser heating of microcapsule cluster

Next, consider the peculiarities of laser radiation absorption by a cluster of closely spaced microcapsules, each being a bilayer particle consisting of a water core and a light-absorbing composite shell. To a certain extent, this imitates the situation when microcapsules first accumulate in the target zones of a body and then are *en masse* exposed to laser pulse providing light-induced destruction of capsule shells and cargo release.

We simulate the photothermal effect in a cluster of capsules using the FEM technique driven by the COMSOL Multiphysics core. To this end, we construct the numerical 2D-model of a heterogeneous medium including an aquatic environment and ten microcapsules arranged in four layers (Figure 6). This is not exactly the case of volume-bounded hollow particles considered in the previous sections. However, the 3D-to-2D dimension reducing considerably facilitates the FEM-calculations of the numerical problem while keeping unchanged all basic features of laser-capsule interaction.

The mutual arrangement of particles is assumed to be at the nodes of the periodic lattice with a pitch $d$ and an interparticle gap $g = d - R$. Meanwhile, to take into account the natural aggregation of particles under the influence of random microfluxes of host liquid, their final coordinates $\tilde{\mathbf{X}}_p$ were randomized and specified by the following expression: $\tilde{\mathbf{X}}_p = \mathbf{X}_p + \tilde{\boldsymbol{\xi}}_p$, where $\mathbf{X}_p$ is the coordinate vector of the *p*-th node of spatial lattice, and $\tilde{\boldsymbol{\xi}}_p$ is a pair of independent random numbers in the range $\left|\tilde{\boldsymbol{\xi}}_p\right| < g/2$ obeying the normal probability distribution with zero mean. Here, we study only cylindrical and rectangular capsules, while in the latter case the microbars also have random rotations around their out-of-plane axis. To simulate a large-scale cluster of capsules, at the vertical boundaries of the computational domain the symmetry conditions of the optical and thermal fields are set in the form of Floquet-periodicity.

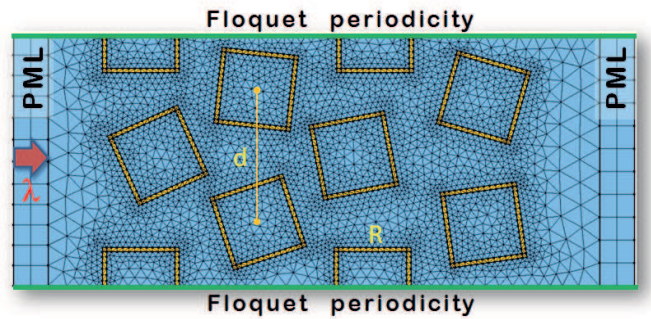

**Figure 6**. COMSOL 2D-model of bar-shaped microcapsule cluster (PML is the perfectly matched layer).

Optical intensity and temperature distributions in the clusters of cylindrical and rectangular microcapsules heated by a 150 μJ laser pulse (λ = 800 nm) are presented in Figures 7(a-d). In the calculations, the pulse length is increased to 1 ns in order to launch the heat transfer in capsule shells and gain the visibility of the temperature distribution. Each pair of pictures with the same capsule shape is plotted for two cases of particle packing density differing in the value of the interparticle gap $g$.

In Figures 7(a, b) one can see a sharp difference between the shell temperatures *T* in cylindrical capsules at different cluster arrangements achieved by the time the laser pulse ends. Indeed, at $g$ = 25 nm the maximum temperature of cylindrical capsules can reach 430 K, while for a more rarefied assembly of capsules ($g$ = 200 nm) their heating is much lower and the maximum temperature of the shells barely reaches approximately 345 K. This is caused by the specific nature of the optical field distribution in closely spaced particles with high light absorption shells ($\alpha^{-1} > h$).

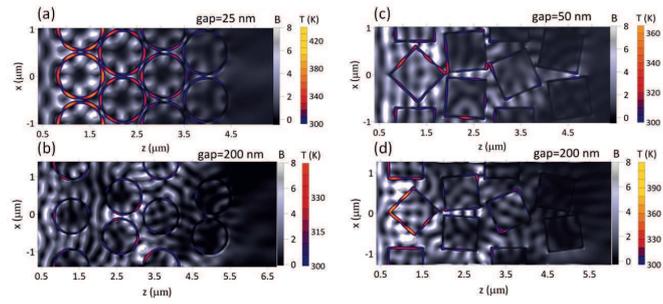

**Figure 7**. Relative intensity *B* (shaded color relief) and temperature *T* (colored contours) distributions in clusters of (a, b) cylindrical and (c, d) bar-shaped microcapsules irradiated by a 150 µJ 1-ns laser pulse and composed into different spatial assemblies: *g* = 25 nm (a), 200 nm (b, d) and 50 nm (c).

When a laser radiation is incident on a cluster of loosely distributed microcapsules (Figure 7b), an intensity pattern of multiple light scattering emerges. Because the scatterers are of mesowavelength dimensions, the ballistic light propagation is damped and the averaged field intensity experiences rapid decrease upon going deeper in the scattering medium. In this case, local optical intensity maxima can appear within the gaps between the capsules, though inside the capsules the light absorption is mainly in the direction of particle illumination by the incoming scattered wave.

Closely spaced cylindrical capsules (Figure 7a) show a rather different distribution of optical field. Here at $g \ll \lambda$, substantially sub-wavelength spaces are formed between neighboring particles. This prevents the optical wave propagation through the illuminated layer of particles, but at the same time causes an increase of laser intensity in the gaps due to near-field amplification effects [27, 28]. In addition, high-absorption shell "locks" the optical wave inside the capsule and aids in a resonant cavity formation that also increases the relative fraction of evanescent leaky modes in the scattered field. As a result, the optical excitation is transmitted to the next layers of cluster not by propagating optical modes, but by coupling surface fields of neighboring capsule resonators [29]. The light-induced heating of deeply embedded particles in this case is more efficient.

Under the same conditions of close packing (Figure 7c), the microcapsules with a cubic cross section are heated by laser pulse considerably weaker. Geometrically, the sharp edges of cubic capsules does not provide for a sufficient number of subwavelength spaces between particles with the ability of extreme intensity increase as the cylindrical capsules do. Thus, the optical absorption regions are localized here mainly in the diffraction maxima of the optical field located inside the shells.

Interestingly, the gap increase between the particles (Figure 7d) does not lead to markedly changes in the temperature field distribution because the cubic shells heating continues to be local. The character of intensity distribution under these conditions resembles the corresponding distribution in a cluster of cylindrical particles. However, it is worthwhile to note that in the sparse packing of cubic capsules a higher temperature is realized in comparison with similar packing of microcylinders (Figure 7b). This is clearly illustrated in Figure 8 that shows the maximum attainable temperature $T_{max}$ in two types of microcapsule clusters considered depending on the gap *g* between particles. Every curve in this figure represents a series averaging for clusters with random particle arrangements (particle positions and rotation angles). For the assessment of a confidence interval of $T_{max}$-evaluation the standard deviation was applied.

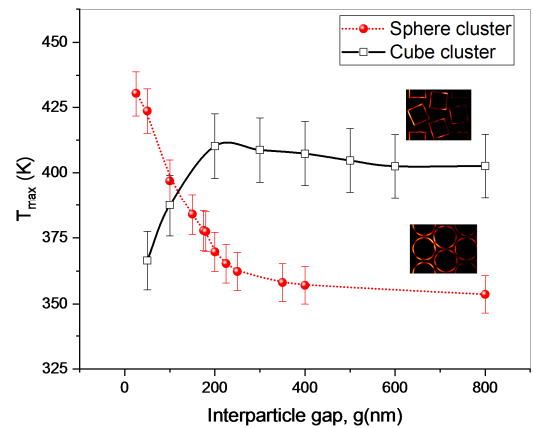

**Figure 8**. Maximal capsule temperature $T_{max}$ for different assembly packing.

As seen, upon changing the packing density of the cluster the capsules of cylindrical and square shapes exhibit different temperature trends. The convergence of cylindrical capsules results in their temperature increase, while closely packed rectangular particles on the contrary are heated up worse than if they are located at a sufficient distance from each other (in sparse cluster). Thus, in order to efficiently heating a cluster of microcapsules of various spatial shapes it is necessary to control the spatial configuration of the particles. This means that cylindrical capsules should be packed as tightly as possible, whereas for square capsules a dense packed assembly is undesirable and the presence of a quarter-wavelength or more space between the particles is necessary.

From the temperature distributions shown in Figures 7(a-d) it follows that in all cases the very first illuminated layer of microcapsules is heated more strongly than other layers of particle assembly. This is due to the peculiarity of optical field propagation through an ensemble of absorbing particles in the multiple scattering regime. The heating wave begins to propagate in the cluster and decays as it moves deeper into the microassembly.

A quantitative assessment of the penetration depth of optical radiation into a cluster of microcapsules, and consequently the degree of their heating can be obtained by constructing the dependence of the mean shells temperature $T_{av}$ on the burial distance $z$ into the microassembly of particles. To this end, we average the 2D-distribution of the capsule temperature $T(x,z)$ in the direction perpendicular to the radiation propagation (along the $x$-axis) according to the formula: $T_{av}(z) = H^{-1} \int T(x,z) dx$, where $H$ is the width of the computational domain. The results of this procedure are shown in Figure 9 as the dependences of mean temperature of the capsules on cluster penetration depth.

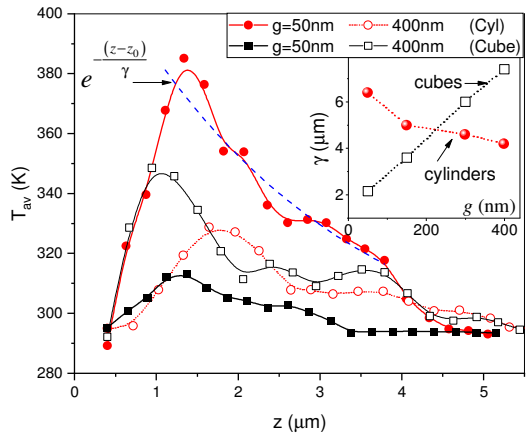

**Figure 9**. Spatially-averaged temperature of microcapsules $T_{av}$ *versus* cluster depth $z$. Inset: Heat penetration depth $\gamma$ for different cluster interparticle gap $g$.

This figure demonstrates that the stationary distribution of mean temperature established in the capsule cluster reaches maximum within the illuminated layer of particles and then decreases. The decaying dynamics can be well fitted by an exponentially decay function: $T_{av}(z) = A + Be^{-(z-z_0)/\gamma}$, where $A$ and $B$ are the fitting coefficients, and $\gamma$ is the characteristic depth of cluster heating. After fitting procedure of the obtained dependences $T_{av}(z)$ the $\gamma$-parameter was calculated, whose dependence $\gamma(g)$ on the interparticle gap in various capsule clusters is shown in the inset to Figure 9.

First of all, one can mention an almost fourfold increase of the heating depth (from 2 µm to almost 8 µm) in the assembly of cubic capsules with the packing density decrease. Recalling maximum particle temperature increase with $g$ (Figure 8), this gives good chances of heating several layers of microcapsules at once which are accumulated in the target region. The heating depth in a cluster of cylindrical capsules is less influenced by particle aggregation degrees, and $\gamma$ is always close to about 5 µm. However, in this case the illuminated layer of core-shell cylindrical capsules demonstrates the most efficient light absorption. This can be used for implementing the regime of a layer-by-layer heating, shell opening and mechanically removing the capsules in a cluster by laser radiation.

## 6. Conclusion

To conclude, the heating by a near-IR laser pulse (800 nm) of a micron-sized two-layer capsules simulating transport microcontainers with water-load and light-absorbing composite shell is theoretically studied. Both single microcapsules and the microassemblies of particles with various spatial shapes and different packing densities are considered. Using FDTD and FEM calculations, the numerical simulations of the optical field distribution inside and near the microcapsules are carried out, and the temporal dynamics of the temperature profiles of microparticles shells is obtained.

Based on the simulation results, one can conclude that the particle morphology introduces specificity in the spatial profile of the optical field and the distribution of light absorption regions. Variation in the geometric shape of a capsule leads to dramatic changes in the distribution of absorbed light energy and, accordingly, particle temperature field. In order to increase the efficiency of absorption of optical radiation in capsule volume and to obtain the maximum heating of absorbing shells, the capsules of cubic, cylindrical and partly rectangular shapes are preferable. Thus, having a 100-fs laser pulse with energy of only 20 µJ it becomes possible to heat a cylindrical microcapsule to the temperature of thermal destruction of its polymeric shell (~ 410 K). Worthwhile notice, the water-filling of a capsule core remains cold. At the same conditions, the ellipsoid-shaped capsule is heated to multiple lower temperatures (~ 340 K).

Aggregation of capsules into microclusters always taking place during mass delivery of payloads to target zones has a significant effect on the absorbing properties of such assemblies. It turns out that when the array of capsules is illuminated by a long laser pulse (1 ns), a quasi-stationary spatial distribution of mean temperature in capsule shells is established. This temperature profile is characterized by an exponential decay when moving deeper into the cluster starting from the illuminated layer of particles. The characteristic depth of cluster heating depends on the shape of individual microcapsules and the distance between them. In closely packed microassemblies, the capsules of cylindrical shape are better heated due to the specific contact heat transfer mechanism. In rarefied capsule cluster, on the contrary, higher shells temperatures are achieved for square-shaped capsules, which due to the presence of sharp edges have the ability of effective utilizing the energy of optical radiation into heat.

## Acknowledgements

This work was partially supported by the Russian Foundation for Basic Research (Grant No. 19-47-700001).

## References

[1] Pavlov A M, Gabriel S A, Sukhorukov G B and Gould D J 2015 *Nanoscale* 7 9686.
[2] Deng L, Li Q, Al-Rehili S, Haneen O, Almalik A, Alshamsan A, Zhang J, and Khashab N M 2016 *ACS Appl. Mater Interfaces* **8** 6859.
[3] A.S. Timin, D. J. Gould, and G.B. Sukhorukov, Expert Opinion on Drug Delivery 2017, DOI: 10.1080/17425247.2017.1285279.
[4] I. Koryakina, D. S. Kuznetsova, D. A. Zuev, V. A. Milichko, A. S. Timin and M. V. Zyuzin, Nanophotonics 2020, 9(1), 39.
[5] F. Ungaro, I. d'Angelo, A. Miro, M. I. La Rotonda, and F. Quaglia, J. Pharm. Pharmacol. 2012, 64(9), 1217.
[6] A.S. Timin, H. Gao, D.V. Voronin, D.A. Gorin, and G.B. Sukhorukov, Adv. Mater. Interfaces 2016, 1600338, DOI:10.1002/admi.201600338.
[7] Y. Ma , X. Liang , S. Tong , G. Bao , Q. Ren , and Z. Dai, Adv. Funct. Mater. 2013, 23, 815.
[8] A.G. Skirtach, A.M. Javier, O. Kreft, K. Köhler, A.P. Alberola, H. Möhwald, W.J. Parak, G.B. Sukhorukov, Angewandte Chemie-International Edition 2006, 45(28), 4612.
[9] A.G. Skirtach, A.A. Antipov, D.G. Shchukin, and G.B. Sukhorukov, Langmuir 2004, 20 (17), 6988.
[10] H. Gao, D. Wen, N.V. Tarakina, J. Liang, A.J. Bushbya, G.B. Sukhorukov, Nanoscale 2016, 8, 5170.
[11] Y.E. Geints, A.A. Zemlyanov, Journal of Applied Physics 2017, 121, 123111, doi: 10.1063/1.4979095.
[12] G.-Y. Liu, C.-J. Chen, D.-D. Li, S.-S. Wang, J. Ji, J. Mater. Chem. 2012, 22, 16865.
[13] S. Link, M.A. El-Sayed, Int. Rev. Phys. Chem. 2000, 19(3), 409.
[14] M. Hu, J. Chen, Z.-Y. Li, L. Au, G.V. Hartland, X. Li, M. Marquez, Y. Xia, Chem. Soc. Rev. 2006, 35, 1084.
[15] Y.E. Geints, E.K. Panina, A.A. Zemlyanov, JQSRT 2019, 236, 106595.
[16] K. Katagiri, M. Nakamura, S. Iseya, A. Matsuda, K. Koumoto, in Proc. Int. Sympos. EcoTopia Sci. ISETS07, 2007, 80.
[17] A. Yashchenok, B. Parakhonskiy, S. Donatan, D. Kohler, A. Skirtach, H. Möhwald, J. Mater. Chem. B 2013, 1, 1223, DOI: 10.1039/c2tb00416j
[18] A.T. Holkem, G.C. Raddatz, G.L. Nunes, A. J. Cichoski, E. Jacob-Lopes, C.R.F. Grosso, C.R. de Menezes, LWT-Food Science and Technology 2016, 71, 302. DOI:10.1016/j.lwt.2016.04.012.
[19] Y.E. Geints, A.A. Zemlyanov, JQSRT 2017, 200, 32.
[20] Y.E. Geints, E.K. Panina, A.A. Zemlyanov, Opt. Quant. Electron. 2018, 50:53. DOI:10.1007/s11082-018-1331-5.
[21] Y.E. Geints, A.A. Zemlyanov, E.K. Panina, Appl. Opt. 2017, 56, 2127.
[22] K.Q. Le, A. Alu, Appl. Phys. Lett. 2014, 105, 141118; DOI:10.1063/1.4898011.
[23] H.S Carslaw, J.C. Jaeger, Conduction of Heat in Solids (2nd ed.), Oxford University Press., UK 1959.
[24] C.F. Bohren and D.R. Huffman, Absorption and Scattering of Light by Small Particles, John Wiley, New York, USA 1983.
[25] M.I. Mishchenko, G. Videen, V.A. Babenko, N.G. Khlebtsov, T. Wriedt, JQSRT 2004, 88, 357.
[26] B. T. Draine and P. J. Flatau, J. Opt. Soc. Am. A 1994, 11, 1491.
[27] A. Mirzaei, A.E. Miroshnichenko, I.V. Shadrivov, and Y.S. Kivshar, Phys. Rev. Lett. 2015, 115, 215501.
[28] C.P. Burrows and W.L. Barnes, Opt. Express 2012, 18(3), 3187.
[29] S. Yang and V. N. Astratov, Appl. Phys. Lett. 2008, 93, 261111. DOI:10.1063/1.2954013.